\documentclass{ws-p8-50x6-00}

\begin{document}

\title{Some results from the NA27 Data}

\author{Wang Shaoshun}

\address{Department of Modern Physics, University of Science and Technology of China, Hefei, 230027, China}

\maketitle

\abstracts{The self-affine analysis and erraticity analysis of 
pseudorapidity gaps are
performed for the data of 400GeV/$c$ pp collisions. The self-affine analysis
has been shown to exhibit a better scaling behavior. 
The self-affine multifractal dimensions
and multifractal spectrum have been obtained. The simulated results using
FRITIOF program can not reproduce the scaling behavior. The analysis of
event-to-event fluctuations has been performed. The increase of event-space
moments $C_{p,q}(M)$ with decreasing phase-space scale is dominated by the
statistical fluctuations. The erraticity analysis based on measuring the
pseudorapidity gaps is also performed.
The entropy-like quantities $S_q$ and $\Sigma_q$ deviate from 1
significantly, implying that both of them are useful to serve as effective
measures of erraticity in multiparticle production. The ln${S_q}$ versus 
$q$ has a quite linear behavior, but the ln${\Sigma _q}$ versus $q$ has
only an approximate linear behavior. The FRITIOF simulated results follow
the same scaling behavior, but the deviations from the
experimental data are rather large.
}

\section{The data}
In the present investigation, the angular distribution of charged particles
produced in pp collisions at 400GeV/$c$ was measured by using the LEBC films
offered by the CERN NA27 collaboration. A total of 3950
non-single-diffractive events ($N\geq $ 4) were measured. The accuracy in
pseudorapidity in the region of interest (-2 $\leq \eta \leq 2)$ is of the
order of 0.1 pseudorapidity units. In order to compare with the experimental
data, we used a Monte-Carlo (MC) 
generator FRITIOF version 7.02 and JETSET 7.3 to
simulate the multiparticle production in 400GeV/$c$ pp collisions. A total
of 4500 non-single-diffractive events ($N\geq \ $4) have been created.

\section{Self-affine analysis}
Since Bia\l as and Peschanski proposed to study nonstatistical fluctuation in
multiparticle production by the method of factorial moments $F_q$($\delta $)
[1], a large variety of experiments were performed to search for the
anomalous scaling behavior
$F_q(\delta )\propto (\delta )^{-\phi _q}\ (\delta \rightarrow 0)$. 
The results have shown that the power-law behavior does not hold exactly for
high energy hadron-hadron collisions [2]. The reason for this is that the
usual procedure for calculating higher-dimensional factorial moments is to
divide phase space into bins with the same $M$ in each direction. This is
called self-similar analysis. However, phase space in high energy
multiparticle production is anisotropic. Wu and Liu have proposed a new
method to calculate the higher-dimensional factorial moments which is called
self-affine analysis [3]. In self-affine analysis, the two-dimensional phase
space region $\Delta \eta \Delta \varphi \ $is divided by $\lambda _\eta
\lambda _\varphi .\ $The shrinking ratios $\lambda _\eta $ and $\lambda
_\varphi \ $are characterized by a parameter
$H={\ln\lambda_\eta}/{\ln\lambda_\varphi}$,
called Hurst exponent, which can be deduced from
the data by fitting two corresponding one-dimensional second-order
factorial-moment saturation curves
\begin{equation}
F_2^{(i)}(M_i)=A_i-B_iM_i^{-C_i}\ (i=\eta ,\varphi) \ ,\quad H_{\eta
\varphi }= (1+C_\varphi )/(1+C_\eta )
\end{equation}

In order to obtain the Hurst exponent from NA27 data, the second order
factorial moment for one-dimensional phase space were calculated and
are fitted to Eq.(3), cf. Table 1.

\vskip 0.1in

\ \ \qquad \quad Table 1 The fitting parameters obtained according to (3).

\begin{center} \begin{tabular}{||c|c|c|c|c||}\hline\hline
Variables & A & B & C & $\chi^2$/NDF \\ \hline
$\eta$ & $1.371\pm 0.025$ & $0.222\pm 0.018$ & $0.425\pm 0.109$ & 8.449/36 \\ \hline
$\varphi$ & $1.509\pm 0.018$ & $0.420\pm 0.039$ & $0.057\pm 0.021$ & 11.16/34 \\ \hline\hline
\end{tabular} \end{center}

From these parameter values, we obtain the Hurst exponent $H_{\eta \varphi
}=0.74\pm 0.07$. The second order two-dimensional factorial
moments are then calculated
with the method of continuously varying scale [4] for 
$H=0.5,0.74,M_\eta =M_\varphi ^H$ ; $H=1.0,2.0$, $M_\varphi =M_\eta ^{1/H}$. 
The results are shown in Fig.1 and are fitted both linearly (dashed 
lines) and quadratically (full lines). The 
parameters of quadratic fit ($y(x)=a+bx+cx^2$) 
are shown in Table.2. We can see that the coefficient 
$b$ which characterizes the strength of 
anomalous scaling is positive only for 
$H=0.74$. The coefficient $c$ of the
quadratic term, characterizing the degree of upward-bending is the smallest
for $H=0.74.$ These results mean that the self-affine analysis exhibits
a better scaling behavior.

\begin{center}
Table 2 The fitting parameters obtained according to (4).

\begin{tabular}{||c|c|c|c||}\hline\hline
$H$ & a & b & c  \\ \hline
0.50 & $0.181\pm 0.007$ & $-0.0213\pm 0.0051$ & $0.0131\pm 0.0015$  \\ \hline
0.74 & $0.144\pm 0.014$ & $0.0243\pm 0.0075$ & $0.0049\pm 0.0014$  \\ \hline
1.00 & $0.194\pm 0.006$ & $-0.0197\pm 0.0031$ & $0.0130\pm 0.0011$  \\ \hline
2.00 & $0.192\pm 0.015$ & $-0.0062\pm 0.0020$ & $0.0100\pm 0.0030$  \\ \hline\hline
\end{tabular}
\end{center}

\vskip 0.2cm

In order to obtain the self-affine multifractal 
spectrum, the two-dimensional factorial moments of continuous order [5] ($q$
from -1 to 4 with step 0.2) are calculated and shown in 
Fig.2. The dotted lines in the figure are from FRITIOF Monte-Carlo,
the dashed lines are from another MC having the
same multiplicity distribution and same number of events as the
experimental data, but no correlations. 
It can be seen that no intermittency behavior can be observed in both 
MC's. 

Using the intermittency exponents $\phi (q)$ for continuous order $q$
obtained by fitting ln$F(q,M)$ versus ln$M$, the
multifractal dimension $D(q)$ and multifractal spectrum $f(\alpha )$ of the
self-affine fractal can be calculated through the following relations 
$$ \tau(q)=q-1-\phi(q),\  D(q)=\tau(q)/(q-1),\  
\alpha =d\tau(q)/dq,\  f(\alpha)=q\alpha -\tau(q). $$
$D(q)$ versus $q$ is shown in Fig.4. $D(q)$ decreases with increasing $q$.
The self-affine multifractal
spectrum $f(\alpha )$ is shown in Fig.5. It is concave downward with a
maximum at $q=0,$ $f(\alpha (0))=D(0)=1.$ These mean that multiparticle
production at 400GeV/$c$ pp collisions is a self-affine multifractal
process.

The index $\mu$ for Levy stable law, defined by the equation:
$\phi (q)/\phi (2)=(q^\mu -q)/(2^\mu -2)$ can be obtained using a method
proposed by Hu Yuan et al. [6].  From Eqs.(5) and (6) we can get the
following relationship [6]
\begin{equation}
1-f(\alpha )\propto (B-\alpha )^{\mu /(\mu -1)},\qquad {\rm for}\quad
\alpha <B. 
\end{equation}

\vskip 0.10in

\noindent
where $B$ is the value of $\alpha $ when $f(\alpha )=1$. From Fig.4, the
value of $\alpha $ for $f(\alpha )=1$ is found to be $B=1.0260.$ The ln(1-$%
f(\alpha ))$ versus ln($B-\alpha )$ are shown in Fig.5.
The slope $C=2.0997\pm 0.0098$ and the Levy index $\mu =C/(C-1)=1.91\pm 0.01$
are obtained through a linear fit.  

\section*{3. The analysis of event-to-event fluctuations}

\indent

The investigation of non-linear phenomena in high energy collisions has
lasted a long time.   Cao and Hwa [7].
proposed to characterize the spacial pattern of an event by using the
horizontally normalized factorial moments. However, their method
is meaningful only for high multiplicity events [8].

When the event multiplicity $N$ is low and the number of bins is high, only
a few events have $n_m\geq q,$ so the statistical fluctuation may be very
large and very little information can be obtained.  In Fig.6 the event space 
moments ln$C_{p,q}(M)$ versus ln$M$ from NA27 data are plotted on the 
left side and ln$C_{p,q}(M)$ versus ln$C_{2,2}(M)$ on the right.
In order to see whether the variation of the ln
$C_{p,q}(M)$ versus ln$M$ is caused by the statistical fluctuations, we
created a MC event sample which has same multiplicity distributions
within $\Delta \eta $ as experimental data, but the particles are randomly
distributed with equal probability. The results 
are shown in Fig.6 as dashed lines. We can see that the
increase of the moments $C_{p,q}(M)$ with decreasing phase-space scale is
indeed dominated by 
the statistical fluctuations.

In order to get the erraticity behavior of the low multiplicity system, a
new method based on measuring the rapidity gaps has been proposed by Hwa and
Zhang [9]. They have proposed the moments 
\begin{equation}
G_q=\frac{1}{N+1}\sum_{i=0}^n x_i^q\ , \ \ \ {\rm and} \ \ \ 
H_q=\frac{1}{N+1}\sum_{i=0}^n(1-x_i)^{-q} \quad
\end{equation}
as measures of spatial patterns in terms of rapidity gaps 
$x_i=X_{i+1}-X_i,\ i=0,\cdots ,N$, 
with $X_0=0$ and $X_{N+1}=1.$ 
From this we can obtain the entropy-like quantities 
\begin{equation}
S_q=\frac{\langle G_q\ln G_q>}{<G_q^{st}\ln G_q^{st}>}{\ ,}  \ \ \ {\rm and} 
\ \ \ \Sigma {}_q=\frac{\langle H_q\ln H_q>}{<H_q^{st}\ln H_q^{st}>}\quad . 
\end{equation}
where $G_q$ and $H_q$ for experimental data, $G_q^{st}$ and $H_q^{st}$ for
pure statistical fluctuations. How much degree of $S_q$ and $\Sigma_q$ deviate
from 1 is a measure of erraticity in multiparticle production based on
rapidity gaps. We calculate the $S_q$ and $\Sigma_q$ only in 
central region (-2 $\leq \eta \leq 2$) and drop the events which 
has less than six particles in
this region. A total of 2515 events for NA27 data and 2723 events for
FRITIOF have been selected.

The results for $S_q$ are shown in Fig.7a. It 
can be seen that $S_q$ deviate from 1 significantly
and the ln$S_q$ versus $q$ has a quite linear behavior. This means that $S_q$
satisfies the exponential relationship $S_q\propto e^{\alpha q}$. 
The straight lines are the linear fit to the experimental data. The fit
parameter is listed in Table 3. This result is different from the power
law behaviour claimed by Hwa and Zhang [9], $S_q\propto q^{\alpha 1}$, 
cf. the linear fit to the 
ln$S_q$ versus ln$q$ plotted in Fig.7b and the 
fitting parameter listed in Table 3. 
The results from FRITIOF MC 
are also shown in Fig.8 (open circles) and Table 3. We can see that
there is a same scaling behavior for the FRITIOF MC,
although the values of $S_q$ deviate from experimental data sufficiently
large.

\begin{center}
Table 3 The fit parameters obtained according to (13) and (14).

\begin{tabular}{||c|c|c|c|c||}\hline\hline
 EVENT SAMPLE & $\alpha $&$\chi ^2/NDF$ &$\alpha 1$ &$\chi ^2/NDF$ \\ \hline
experimental data & $0.133\pm 0.002$& 0.42 & $0.66\pm 0.06$ & 18.8   \\ \hline
FRITIOF & $0.078\pm 0.002$& 0.92 & $0.38\pm 0.04$ & 10.2    \\ \hline
\end{tabular}
\end{center}

\vskip0.5cm

The results for $\Sigma_q$ are shown in Fig.8. It can be seen that
$\Sigma_q$ deviate from 1 significantly, but ln$\Sigma_q$ versus $q$
only has an approximately linear behavior
$\Sigma {}_q\propto e^{\beta q}$,
when $q\geq 2$. For the experimental data, $\beta =$ 0.30$\pm 0.04.$ For the
FRITIOF Monte-Carlo event sample, $\beta =$ 0.15$\pm $0.03 . 

The value of $\beta$ (0.15$\pm $0.03)
for the FRITIOF MC is much smaller than that ($\beta =$ 0.30$\pm 0.04$)
for the experimental data. The $S_q$ and $\Sigma _q$
deviate from 1 significantly implies that both of them are useful to serve
as effective measures of erraticity in multiparticle production.

\section*{Acknowledgements}
We are grateful to the CERN NA27 Collaboration for offering the LEBC films.

\noindent
Project 19975045 is supported by the National Natural Science Foundation of
China.

\section*{References}

\ \ \ \ \ [1] A. Bia\l as, R. Peschanski, Nucl. Phys. B 273 (1986) 703 ; B308
(1988) 

\hskip 0.5cm 857.

[2] E. A. De Wolf, I. M. Dremin, W. Kittel, Phys. Rep. 270 (1996) 1.

[3] Wu Yuanfang, Liu Lianshou, Phys. Rev. Lett. 70 (1993) 3197.

[4] Liu Lianshou et. al., Z. Phys.C 69 (1996) 323.

[5] R. C. Hwa, Phys. Rev. D51 (1995) 3323.

\quad Zhang Jie, Wang Shaoshun, Phys. Rev. D55 (1997) 1257.

[6] Hu Yuan et. al.,Chin. Phys. Lett. 16 (1999) 553.

[7] Z. Cao, R. C. Hwa, Phys. Rev. Lett. 75 (1995) 1268;

\quad Phys. Rev. D 53 (1996) 6608; Phys. Rev. D 54 (1996) 6674.

[8] J. Fu et. al., Phys. Lett. B 472 (2000) 161.

[9] R. C. Hwa, Q.-H. Zhang, Phys. Rev. D 62 (2000) 0140003.
\vskip1cm

\section*{Figure Captions}

\hskip0.65cm
Fig.1. ln$F_2$ versus ln($M_\eta M_\varphi )$ for different values of
the Hurst-exponent.
\vskip0.1cm

Fig.2 \ ln$F_q$ versus ln$M$ for continuous order $q$.
\vskip0.1cm

Fig.3 \ $D(q)$ versus $q$.
\vskip0.1cm

Fig.4 \ $f(\alpha)$ versus $\alpha$.
\vskip0.1cm

Fig.5 \ ln$(1-f(\alpha))$ versus ln$(B-\alpha)$.
\vskip0.1cm

Fig.6 \ The event space moments ln$C_{p,q}(M)$ versus ln$M$.
\vskip0.1cm

Fig.7 \ ln$S_q$ versus $q$ and ln$q$.
\vskip0.1cm

Fig.8 \ ln$\Sigma_q$ versus $q$.

\end{document}